\documentclass[aps,prd,amsfonts,nofootinbib,supercriptaddress]{revtex4}
\usepackage{amssymb,amsmath}
\usepackage{graphicx}
\usepackage{epsfig}
\usepackage{bbm}

\usepackage[T1]{fontenc}
\usepackage[latin9]{inputenc}
\usepackage{amssymb}
\usepackage{esint}

\newcommand{\beq}{\begin{equation}}
\newcommand{\eeq}{\end{equation}}
\newcommand{\bea}{\begin{eqnarray}}
\newcommand{\eea}{\end{eqnarray}}
\newcommand{\vc}[1]{{\textbf{#1}}}

\newcommand{\mc}[1]{\mathcal{#1}}

\newcommand{\noun}[1]{\textsc{#1}}

\begin{document}

\title{On the divergences of inflationary superhorizon perturbations}

\author{K.~Enqvist$^{1,2}$}
\author{S.~Nurmi$^{1,2}$}
\author{D.~Podolsky$^2$}
\author{G.~I.~Rigopoulos$^2$}

\affiliation{$^1$Physics, Department, University of Helsinki;
$^2$Helsinki Institute of Physics, P.O.Box 64, FIN-00014, University
of Helsinki, Finland}

\begin{abstract}

\noindent We discuss the infrared divergences that appear to plague
cosmological perturbation theory. We show that within the stochastic
framework they are regulated by eternal inflation so that the theory
predicts finite fluctuations. Using the $\Delta N$ formalism to one
loop, we demonstrate that the infrared modes can be absorbed into
additive constants and the coefficients of the diagrammatic
expansion for the connected parts of two and three-point functions of the
curvature perturbation. As a result, the use of any infrared cutoff
below the scale of eternal inflation is permitted, provided that the
background fields are appropriately redefined. The natural choice
for the infrared cutoff would of course be the present horizon;
other choices manifest themselves in the running of the correlators.
We also demonstrate that it is possible to define observables that are
renormalization group invariant. As an example, we derive a non-perturbative,
infrared finite and renormalization point independent
relation between the two-point correlators of the curvature perturbation
for the case of the free single field.

\end{abstract}

\maketitle

\section{Introduction}

Primordial perturbations generated during inflation are conveniently
characterized by the gauge-invariant curvature perturbation which
can be directly related to the observed CMB anisotropies
\cite{WMAP3}. The study of the properties of the curvature
perturbation arising in various inflationary models has led to a
remarkable confrontation of early universe theory with cosmological
observations. The increasing observational precision has placed
increasing demands on the theory and over the past few years results
beyond linear theory have been sought, mainly focusing, but not
limited to, non-gaussianity. It is well known however that the
$n$-point correlators of the curvature perturbation contain IR
divergences as a consequence of the approximative scale invariance
and gaussianity of inflationary perturbations\footnote{More
accurately, the expressions depend on an IR cutoff L.}. Although
correlators extending to superhorizon scales as such are not
observable, the infrared behavior can manifest itself non-trivially
both in classical and quantum field theory when computing higher
order perturbative corrections. Moreover, in principle we only have
to wait long enough, and the infrared modes will become observable.
Therefore, a satisfactory understanding of the theory requires that
these IR divergences can be controlled.

To demonstrate the problems related to the infrared divergences, it
is enough to consider two point correlators and single field
inflation. Using the $\Delta N$ formalism \cite{DeltaN}, the curvature
perturbation $\zeta({\bf x})$ can be expressed in a simple form
  \beq
  \label{zeta}
  \zeta({\bf x})=N'\phi({\bf x})+\frac{1}{2}N''(\phi^2({\bf
  x})-\langle\phi^2\rangle)+\ldots\ ,
  \eeq
where $N$ is the number of e-folds and $\phi$ denotes perturbations
of the inflaton field on spatially flat slices. The first order
contribution to the two point correlator of the curvature
perturbation behaves as
  \beq
  \label{first}
  \langle\zeta({\bf x}_1)\zeta({\bf x}_2)\rangle_{(1)}=\left(N'\right)^2G(\left|{\bf x}_1-{\bf x}_2\right|)
  \sim
  -(N')^2\mc{P}_{\phi}{\rm ln}\Big(\frac{|{\bf x}_1-{\bf x}_2|}{L}\Big)\ ,
  \eeq
where $G(\left|{\bf x}_1-{\bf x}_2\right|)=\langle\phi({\bf x}_1)\phi({\bf x}_2)\rangle$ is the correlation function of the scalar field,
$\mc{P}$ gives the amplitude of the inflaton perturbations
$\phi$, and we have considered the case of scale invariant fluctuations. The scale $L$ is an a priori arbitrary cutoff needed to
regulate the infrared divergences of the two point correlator. In reality, the theory may contain a physical scale beyond which
scale invariance is broken and $L$ may be related to that scale. For instance, one could argue that in our local patch
inflation has lasted only a finite time so that beyond some infrared scale, the assumption of scale invariance
necessarily breaks down. In any case, in the actual experiments only differences of correlators
are measured and they remain finite even in the limit
$L\rightarrow\infty\ ,$
  \beq
  \label{difference}
  \langle\zeta({\bf x}_1)\zeta({\bf x}_2)\rangle_{(1)}-\langle\zeta({\bf x}_1)\zeta({\bf x}_3)\rangle_{(1)}
  \sim
  -(N')^2\mc{P}_{\phi}{\rm ln}\Big|\frac{{\bf x}_1-{\bf x}_2}{{\bf x}_1-{\bf
  x}_3}\Big|\ .
  \eeq
The value of the IR cutoff $L$ is therefore irrelevant when
considering observable quantities calculated using first order
perturbation theory.

The situation becomes more complicated beyond the linear order since the divergent part of the correlators
will no longer be simply an additive constant. For example, the second order contribution to the two point correlator behaves as
  \beq
  \label{second}
  \langle\zeta({\bf x}_1)\zeta({\bf x}_2)\rangle_{(2)}
  \sim
  -N'N'''\langle\phi^2\rangle\mc{P}_{\phi}{\rm ln}\Big(\frac{|{\bf x}_1-{\bf x}_2|}{L}\Big)+
  \frac{1}{2}(N'')^2\mc{P}^2{\rm ln}^2\Big(\frac{|{\bf x}_1-{\bf x}_2|}{L}\Big)\ ,
  \eeq
which in the limit $L\rightarrow\infty$ contains divergences
depending also on the separation of the two points ${\bf x}_1$ and
${\bf x}_2$.

There are two distinct issues associated with the appearance of these IR divergences. The first is
whether $G(\left|{\bf x}_1-{\bf x}_2\right|)$ is actually IR divergent. The divergence is the outcome
of the assumption that the spectrum is scale invariant, an assumption that will probably break down
after some scale L. One might expect that the complete theory of inflation predicts a well defined
value for $G(\left|{\bf x}_1-{\bf x}_2\right|)$, maybe large but free of divergences. The other issue
is the relation of this (large) value of $G(\left|{\bf x}_1-{\bf x}_2\right|)$ to the measurements of
an observer accessing only a small patch of the whole universe. In other words: how do the predictions
of the theory defined on the largest possible scales translate into predictions for measurements on much smaller scales?
These problems have been discussed in the literature \cite{Lyth1,Boubekeur:2005fj,Lyth:2006gd, bartoloetal, riotto-sloth},
but there seems to be no agreement on the correct approach.

In this paper we address both issues. First, we show that the
variance of the field $\langle\phi^2\rangle$ is finite and well
defined for generic slow-roll inflation. We prove this using the
framework of stochastic inflation which accurately describes the
fluctuations of quantum fields on superhorizon scales. Furthermore,
we note that since IR fluctuations are only observationally relevant
in regions which have exited the eternally inflating regime and have
eventually thermalized, the scale of eternal inflation $V(\phi_{{\rm
EI}}) \lesssim \epsilon(\phi_{{\rm EI}})M_{P}^{4}$ provides an
ultimate upper limit for observable fluctuations.
Only below the length scale $L_{\rm EI}$, corresponding to
$\phi_{\rm EI}$, does the notion of a well defined evolving
background field apply and the spacetime can be
approximated by a global FRW metric. We then proceed to consider how
predictions at this scale can be related to predictions for
observations restricted in patches of size $M < L < L_{\rm EI}$ by
applying a systematic renormalization prescription. In particular,
we focus on the renormalization of the classical one loop
expressions for the 2-point and 3-point functions of the curvature
perturbation. We find that the renormalization amounts to shifting
the background fields in the connected parts of the correlators. Furthermore,
we note that apart from some contrived models, the exact position of the patch of size
$M$ in the larger universe is irrelevant in single field inflation,
to the extent that the background evolution corresponding to the
observable patch $M$ has been defined. Therefore, averaging over all
possible embeddings of $M$ in the whole universe has no meaning in
this context.

The paper is organized as follows: In sections II and III we use the
formalism of stochastic inflation to demonstrate that for potentials
supporting slow-roll inflation, the one point functions of the
scalar field and the curvature perturbation are finite as long as
the energies are sub-planckian. Therefore there are no real IR
divergences associated with inflationary fluctuations. Then, in
section IV we relate predictions at scales $L$ to scales $M < L$
using the $\Delta N$ formalism, and argue that a renormalization
group equation can be used to relate predictions obtained using two
different cutoffs/renormalization scales. In section V we discuss a concrete
example, the free field, and demonstrate that it is possible to
construct an observable that is both infrared finite and renormalization
point independent. In section VI we close with a
discussion and a comparison of our findings to recent work on the
subject.

\section{Eternal inflation and one-point correlation functions of the inflaton}

In this Section, we will show how effects of eternal inflation lead
to the regularization of the infrared divergences present in the correlation
functions of the inflaton. Although we will focus on the chaotic
inflation scenario \cite{StarobinskyStochasticInflation,StarobinskyYokoyama}, our analysis can be easily extended to the case
of new inflation \cite{StarobinskyStochasticInflation,AlStarOthers}, inflation driven by multiple
fields \cite{SalopekBondOthers} or to the case of the curvaton scenario \cite{curvaton}.

Let us suppose that primordial inflation is driven by a single scalar
field $\phi$ with the potential $V(\phi)$ satisfying the slow roll initial
conditions $\epsilon=\frac{1}{2}{M_{P}^{2}}\left({V_{,\phi}}/{V}\right)^{2}\ll1$,
$\eta=M_{P}^{2}{V_{,\phi\phi}}/{V}\ll1$. During one Hubble time
$\Delta t\sim H^{-1}$ the value of the inflaton field changes by
\begin{equation}
\Delta\phi\sim\dot{\phi}\Delta t\sim\frac{V_{,\phi}}{3H^{2}}\sim\frac{M_{P}^{2}}{8\pi}\frac{V_{,\phi}}{V}.
\label{eq:Deltaphiclass}
\end{equation}
On the other hand, at the same time scale fluctuations of the inflaton
$\delta\phi$ are generated with a characteristic wavelength $l\sim k^{-1}\sim H^{-1}$.
These fluctuations have a randomly distributed amplitude among different
causally disconnected regions (Hubble patches), and the width of this
distribution is given by
\begin{equation}
|\delta\phi|\sim\frac{H}{2\pi}\sim\sqrt{\frac{2V}{3\pi M_{P}^{2}}}.
\label{eq:Deltaphiquant}
\end{equation}
As a result, observers living in different Hubble patches see different
expectation values of the inflaton, because  for them long wavelength (superhorizon)
fluctuations $\delta\phi$ are physically indistinguishable from the
inflaton zero mode.

In some Hubble patches the sign of the infrared fluctuation $\delta\phi$
is positive, in others it is negative and therefore may overcome the
effect of the classical force (\ref{eq:Deltaphiclass}) acting on
the inflaton. When $\Delta\phi\lesssim|\delta\phi|$, stochastic fluctuation
$\delta\phi$ may lead to the effective growth of the of expectation
value of the inflaton $\phi$ in a given Hubble volume. This regime
is denoted as \emph{eternal inflation} because there will be always
Hubble patches where the expectation value of the inflaton continues
to grow \cite{LindeEternal}.

One can see that the stochastic force becomes more important than
the classical one when $|(M_{P}^{2}/V)V_{,\phi}|\lesssim\sqrt{V/M_{P}^{2}}$
or
\begin{equation}
V(\phi_{{\rm EI}})\gtrsim\epsilon(\phi_{{\rm EI}})M_{P}^{4},
\label{eq:EternalInflationBoundary}
\end{equation}
where $\epsilon$ is the slow roll parameter. For example, in the
$\lambda\phi^{4}$ chaotic inflation model the condition (\ref{eq:EternalInflationBoundary})
is satisfied when $\phi>\phi_{{\rm EI}}=\lambda^{-1/6}M_{P}$. Since
the upper bound on possible values of field corresponds to the Planckian
energy density, we conclude that the regime of eternal inflation is
realized at $\lambda^{-1/6}M_{P}<\phi<\lambda^{-1/4}M_{P}$, a very
wide interval of possible values of the inflaton, since $\lambda\sim10^{-13}$.

Larger values of field correspond to larger scales of perturbations
in the chaotic inflationary scenario, and one may conclude that the
regime of eternal inflation can influence the infrared behavior of
the correlation functions of both $\phi$ and the curvature perturbation
$\zeta$. In fact, as Woodard has shown \cite{Woodard}, leading
infrared logarithmic divergences in the correlation functions of $\phi$
and $\zeta$ can be reproduced in the formalism of stochastic inflation
\cite{StarobinskyStochasticInflation} which is thus well suited
for treating the regime of eternal inflation.

In this formalism one decomposes the Heisenberg operator for the field
$\phi$ into the infrared and ultraviolet parts according to the prescription
\begin{equation}
\hat{\phi}(N,{\bf x})=\Phi(N,{\bf x})+\frac{1}{(2\pi)^{3/2}}\int d^{3}k\,\theta(k-\delta aH)\left(\hat{a}_{k}\phi_{k}(N)e^{-i{\bf kx}}+\hat{a}_{k}^{\dagger}\phi_{k}^{*}(N)e^{i{\bf kx}}\right)+\delta\phi,
\label{eq:IRUVdec}
\end{equation}
where $N=\log\, a$ is the number of inflationary e-folds, $\delta\ll 1$
is a small parameter%
\footnote{This parameter is intrinsically related to the decoherence scale $l_{{\rm d}}$,
where perturbations which are leaving the horizon finally become classical
(i.e., to the scale, where decaying mode can be neglected as compared
to the growing one). The scale decoherence $l_{{\rm d}}$ corresponds
to few e-folds after perturbations leave the horizon. Note that
$\delta$ cannot be made arbitrarily small as it is related to the slow
roll parameters \cite{StarobinskyStochasticInflation}.%
}, $\theta(x)$ is the Heaviside step function and $\delta\phi$ is
a small contribution suppressed by additional powers of the slow roll
parameters. We are especially interested in the dynamics of the infrared
part $\Phi$ of the inflaton field, which is responsible for
the infrared divergences.

Substituting the decomposition $(\ref{eq:IRUVdec})$ into the equation
of motion for the Heisenberg operator $\hat{\phi}$, one can show
that the infrared part $\Phi$ satisfies the Langevin equation
\begin{equation}
\frac{\partial\Phi}{\partial N}=-\frac{1}{3H^{2}}V_{,\Phi}+\frac{f}{H},\label{eq:Langevin}
\end{equation}
where $f$ is the composite operator containing contributions of the
ultraviolet modes and having the correlation property
\begin{equation}
\langle f(N)f(N')\rangle=\frac{H^{4}}{4\pi^{2}}\delta(N-N'),
\end{equation}
as can be checked by a direct calculation of this correlator in
the Bunch-Davies vacuum state. The remarkable property of the equation
(\ref{eq:Langevin}) is that all its terms commute with each other
and therefore can be considered as classical stochastic quantities. The Langevin equation (\ref{eq:Langevin}) describes
the process of the inflaton random walk due to the effect of generated
superhorizon fluctuations.

At the next step, by the Stratonovich prescription one proceeds from
the Langevin equation (\ref{eq:Langevin}) to the Fokker-Planck equation
\begin{equation}
\frac{\partial P}{\partial N}=-\frac{1}{3\pi M_{P}^{2}}\frac{\partial^{2}}{\partial\phi^{2}}(VP)+
\frac{M_{P}^{2}}{8\pi}\frac{\partial}{\partial\phi}\left(V^{-1}\frac{\partial V}{\partial\phi}P\right),\label{eq:FP}
\end{equation}
which governs the evolution of the probability $P(\phi,N)$ to measure
a given expectation value of the inflation in a given Hubble patch.
Its solution can be expressed in the form
\begin{equation}
P(\phi,N)=v(\phi)e^{-v(\phi)}\sum_{n=0}^{+\infty}c_{n}\psi_{n}(\phi)e^{-E_{n}N},
\label{eq:FPsolution}
\end{equation}
where
\begin{equation}
v(\phi)=\frac{3M_{P}^{4}}{16V(\phi)},
\label{eq:v}
\end{equation}
while $\psi_{n}$ and $E_{n}$ are the eigenfunctions and eigenvalues
of the  Schr\"{o}dinger equation given by
\begin{equation}
-\frac{1}{2}\psi_{n}^{''}+\frac{1}{2}(-v^{''}+(v^{'})^{2})\psi_{n}=E_{n}\psi_{n}\frac{8\pi v}{M_{P}^{2}}.
\label{eq:Schr}
\end{equation}
Due to the supersymmetric form of the potential in the Schr\"{o}dinger
equation (\ref{eq:Schr}) all eigenvalues satisfy the condition $E_{n}\ge0$, and it
is trivial to see that the ground state corresponds to a zero eigenvalue.
To be precise, the ground state corresponds to $E_{0}=0$ if and only
if the corresponding eigenmode is normalizable. However, this is always
the case for chaotic inflationary models with $V(\phi)$ bounded from below and growing as
$\phi\to0$.

The Fokker-Planck equation (\ref{eq:FP}) should be supplemented with
initial conditions (in particular, they define constants $c_{n}$
with $n\ge1$ in the solution (\ref{eq:FPsolution})). It is physically
reasonable to choose them near the hypersurface of Planckian energy
density $V(\phi_{{\rm max}})=M_{P}^{4}$ corresponding to the cosmological
singularity.%
\footnote{Contrary to the general lore, the regime of eternal inflation does
not solve the problem of cosmological singularity, and the latter
remains unavoidable even in eternally inflating space-time \cite{VilenkinSingularities}.
By choosing such initial conditions for the Fokker-Planck equation,
we neglect Hubble patches where the singularity is achieved but the
overall energy density in the patch is dominated by matter fields
other than the inflaton field $\phi$.%
} If the universe started its evolution from the single Hubble patch
of the nearly Planckian size (with, say, $H_{i}^{2}\sim\frac{1}{2}M_{P}^{2}$
), one takes
\begin{equation}
P(\phi,N_{i})=\delta\left(\phi-\frac{1}{\sqrt{2}}\phi_{{\rm \max}}\right),
\label{eq:InitialCond}
\end{equation}
where $N_{i}=\log\, a_{i}$ and $a_{i}$ is the initial value of the
scale factor in that Hubble patch.%
Observe that one cannot choose initial conditions of the form $P_{i}=\delta(\phi-\phi_{{\rm max}})$
due to breakdown of the quasiclassical approximation used for the
derivation of the Langevin equation (\ref{eq:Langevin}).

From the solution (\ref{eq:FPsolution}) we see that in the deep infrared
limit $N=\log\, a\,\to\,\infty$ or, more precisely, at $\log\, a \gg E_1^{-1} $ only the term with $E_{0}=0$ survives
in the solution (\ref{eq:FPsolution}), i.e., the distribution function
$P(\phi,N)$ reaches its time-independent asymptotics
\begin{equation}
P_{{\rm LV}}(\phi)\sim V(\phi)^{-1}\exp\left(-\frac{3M_{P}^{4}}{8V(\phi)}\right)
\label{eq:LindeVilenkin}
\end{equation}
which is called the Linde-Vilenkin wavefunction of the Universe \cite{LindeWaveFunction,VilenkinWaveFunction}.
Therefore, one-point correlation functions of the inflaton field are
given in the limit $N\to\infty$ by the expression
\begin{equation}
\langle\phi^{n}\rangle_{{\rm IR}}=\frac{1}{{\rm N}}\int_{\phi_{{\rm min}}}^{\phi_{{\rm max}}}d\phi\phi^{n}P_{{\rm LV}}(\phi),
\label{eq:PhiCorrelators}
\end{equation}
where $\phi_{{\rm min}}$ is given by the condition $\epsilon(\phi_{{\rm min}})\sim1$,
the normalization is defined as ${\rm N}=\int_{\phi_{{\rm min}}}^{\phi_{{\rm max}}}d\phi\, P(\phi)$,
and $n$ is arbitrary (limited only by the condition that the corresponding
integrals converge).

Therefore, we have explicitly shown that the naively diverging infrared
parts of all one-point correlation functions of the inflaton field
are rendered finite once one takes the effects of eternal inflation into account.

One may observe that, although being finite, the correlators $\langle\phi^{n}\rangle_{{\rm IR}}$
still remain very large. As an example, one can take chaotic inflationary
model with the potential $V(\phi)=\frac{1}{2}m^{2}\phi^{2}$. Using
Eqs. (\ref{eq:LindeVilenkin}), (\ref{eq:PhiCorrelators})
and taking into account the the upper bound for the value of the inflaton
field is $\phi_{{\rm max}}=\sqrt{2}{M_{p}^{2}}/{m}$, we find
that $\langle\phi^{2}\rangle_{{\rm IR}}\approx0.96{M_{P}^{4}}/{m^{2}}$
and $\langle\phi^{4}\rangle_{{\rm IR}}\approx0.15{M_{P}^{8}}/{m^{4}}$.
The level of non-gaussianity generated by eternal inflation
$\langle\phi^{4}\rangle_{{\rm IR}}-3\langle\phi^{2}\rangle_{{\rm IR}}^{2}\approx-2.61{M_{P}^{8}}/{m^{4}}$
is extremely large. However, this amount of non-gaussianity
will never be accessible for an observer within a given Hubble patch
since the horizon of this Hubble patch will never cross the hypersurface
of eternal inflation \cite{LindeBook}.

For completeness, let us discuss what happens with the Fokker-Planck
probability distribution if  eternal inflation comes to
an end before the stationary asymptotics (\ref{eq:LindeVilenkin})
is reached.

At $\phi<\phi_{{\rm EI}}$ one has to neglect the first, stochastic,
term in the right-hand side of the Fokker-Planck equation (\ref{eq:FP}),
and the latter acquires the form
\begin{equation}
\frac{\partial P}{\partial N}=\frac{M_{P}^{2}}{8\pi}\frac{\partial}{\partial\phi}\left(\frac{1}{V}\frac{\partial V}{\partial\phi}P\right).
\label{eq:FPmodified}
\end{equation}
Its solution is given by
\begin{equation}
P(\phi,N)=\frac{V}{V_{,\phi}}f\left(N+\int_{\phi_{{\rm min}}}^{\phi}d\phi\frac{8\pi}{M_{P}^{2}}V\left(\frac{\partial V}{\partial\phi}\right)^{-1}\right),
\label{eq:ProbabilityTransported}
\end{equation}
where the function $f$ should be determined from the evolution of
the probability distribution $P$ in the regime of eternal inflation.
The physical meaning of the solution (\ref{eq:ProbabilityTransported})
is that the probability is transported unchanged along the hypersurface
\begin{equation}
N+\int_{\phi_{{\rm min}}}^{\phi}d\phi\,\frac{8\pi}{M_{P}^{2}}V\left(\frac{\partial V}{\partial\phi}\right)^{-1}=N_{{\rm total}}.
\end{equation}
Note that we count the number of e-folds starting from the
beginning of inflation instead of from its end, i.e., we define $N=\log\,\left({a}/{a_{i}}\right)$,
where $a_{i}$ is the initial value of the scale factor.

Finally, we note that the multipoint correlation functions of the form $\langle \phi (x_1) \phi (x_2) \cdots \phi (x_n) \rangle$, where $x_1$, $x_2$, $\ldots$, $x_n$ are separated by distances larger than the coarse graining scale (i.e., the horizon scale) depend on the Starobinsky's decoherence parameter $\delta$ and cannot be calculated without considering carefully what happens at near-horizon scale.

\section{one-point correlation functions of the curvature perturbation}

Let us now show that one-point correlation function of the curvature
perturbation $\langle\zeta^{n}\rangle$ also remains finite if one
takes properly into account the effects of eternal inflation. Our
goal will be to calculate the average duration of the inflationary
stage $\langle N_{{\rm total}}\rangle$ as well as its higher order
correlation functions $\langle N_{{\rm total}}^{n}\rangle$ for
arbitrary $n$. Correlation functions of the curvature perturbation
$\zeta$ at the end of inflation (i.e., at $N=N_{{\rm total}}$) are
trivially related to the latter:
\begin{equation}
\langle\zeta^{n}(N_{{\rm total}})\rangle=\langle(N_{{\rm total}}-\langle N_{{\rm total}}\rangle)^{n}\rangle.
\label{eq:zetatotal}
\end{equation}
To calculate these one-point correlation functions, we will use the
method developed by  Starobinsky in \cite{StarobinskyStochasticInflation}
and expand it for the case of chaotic infationary scenario.

It is basically guaranteed that inflation will come to the end in
a given Hubble patch when the inflaton expectation value drops below
the boundary $\phi_{{\rm EI}}$ defined by the Eq. (\ref{eq:EternalInflationBoundary}),
and the evolution of the inflaton field becomes deterministic. However,
there are still stochastic fluctuations of the inflaton field which
may lead to a sudden change of $\phi$ in a given Hubble patch from
a value $\phi>\phi_{{\rm EI}}$ to the value $\phi_{{\rm min}}<\phi\ll\phi_{{\rm EI}}$.
Thus, to find the total expected number of e-folds for the given
Hubble patch one has to use the formalism of stochastic inflation.

The probability distribution for the end
of inflation or, in other words, the total number of e-folds $N_{{\rm total}}$
can be determined from the probability distribution $P(\phi,N)$ from conservation of probability by
\begin{equation}
w(N_{{\rm total}})=P(\phi,N)\left|\left(\frac{\partial\phi}{\partial N_{{\rm total}}}\right)_{N}\right|
=\frac{M_{P}^{2}}{8\pi}\lim_{\phi\to\phi_{{\rm min}}}\left|\frac{1}{V}\frac{\partial V}{\partial\phi}\right|\, P(\phi,N_{{\rm total}}).
\label{eq:w}
\end{equation}
Instead of dealing with this distribution function directly, we will
calculate the correlation functions
\begin{equation}
Q_{n}(\phi)=\int_{N_{i}}^{+\infty}dN\, N^{n}P(\phi,N),
\end{equation}
where $N_{i}=\log\, a_{i}$ and $a_{i}$ is the value of the scale factor
at the beginning of inflation. The moments $Q_{n}$ can be naturally
related to the correlation functions of the total number of e-folds
$\langle N_{{\rm total}}^{n}\rangle$ and curvature perturbation correlation
functions $\langle\zeta^{n}\rangle$ , as we will see later.

The zero moment $Q_{0}(\phi)$ can be determined from the equation
\begin{equation}
-\frac{1}{3M_{P}^{2}}\frac{\partial^{2}}{\partial\phi^{2}}(VQ_{0})+\frac{M_{P}^{2}}{8\pi}
\frac{\partial}{\partial\phi}\left(\frac{1}{V^{2}}\frac{\partial V}{\partial\phi}(VQ_{0})\right)=-P(\phi,N_{i}).
\label{eq:Q0}
\end{equation}
After taking into account that $Q_{0}(\phi=\phi_{\max})=0$, i.e.,
that the probability flow through the hypersurface of Planckian energy
density is absent, one finds
\begin{equation}
V(\phi)Q_{0}(\phi)=3M_{P}^{2}e^{-2v(\phi)}\int_{\phi}^{\phi_{{\rm max}}}d\phi'\, e^{2v(\phi')}\int_{\phi_{{\rm min}}}^{\phi'}d\phi''\,
P(\phi,N_{{\rm i}}),
\label{eq:Q0solution}
\end{equation}
where $v(\phi)$ is again given by Eq. (\ref{eq:v}). From
this expression one can immediately see that the probability distribution
function (\ref{eq:w}) for the total duration of inflationary stage
is properly normalized:
\begin{equation}
\frac{M_{P}^{2}}{8\pi}\lim_{\phi\to\phi_{{\rm min}}}\left|\frac{1}{V}\frac{\partial V}{\partial\phi}\right|Q_{0}=\int_{N_{i}}^{\infty}dN\, w(N)=1.
\end{equation}
In turn, the higher moments $Q_{n}$ are governed by the recursive set
of equations
\begin{equation}
-\frac{1}{3M_{P}^{2}}\frac{\partial^{2}}{\partial\phi^{2}}(VQ_{n})+\frac{M_{P}^{2}}{8\pi}
\frac{\partial}{\partial\phi}\left(\frac{1}{V^{2}}\frac{\partial V}{\partial\phi}(VQ_{n})\right)=-nQ_{n-1},
\label{eq:Qn}
\end{equation}
relating higher moments $Q_{n}$ with the lower ones $Q_{n-1}.$ The
general solution of Eq. (\ref{eq:Qn}) can be written in the form
similar to (\ref{eq:Q0solution}):
\begin{equation}
V(\phi)Q_{n}(\phi)=3M_{P}^{2}e^{-2v(\phi)}
\int_{\phi}^{\phi_{{\rm max}}}d\phi'\, e^{2v(\phi')}\int_{\phi_{{\rm min}}}^{\phi'}d\phi''\, nQ_{n-1}(\phi'').
\label{eq:Qnsolution}
\end{equation}
We immediately conclude that all the higher moments $Q_{n}(\phi)$
are finite, since $Q_{0}(\phi)$ and all the integrals in this expression
are well behaved.

The correlation functions of the total number of e-folds are related
with moments $Q_{n}(\phi)$ according to the prescription
\begin{equation}
\langle N_{{\rm {\rm total}}}^{n}\rangle=\frac{M_{P}^{2}}{8\pi}\lim_{\phi\to\phi_{{\rm min}}}
\left|\frac{1}{V}\frac{\partial V}{\partial\phi}\right|Q_{n}=n\int_{\phi_{{\rm min}}}^{\phi_{{\rm max}}}d\phi\, Q_{n-1}(\phi),
\label{eq:NtotalCorrF}
\end{equation}
where we have used  Eq. (\ref{eq:Qn}).%
\footnote{In the deterministic inflationary regime the first term on the right
hand side can be neglected with respect to the second one (see the
end of the previous Section).} In particular, the average duration of the inflationary stage is given
by
\begin{equation}
\langle N_{{\rm total}}\rangle=\int_{\phi_{{\rm min}}}^{\phi_{{\rm max}}}d\phi\, Q_{0}(\phi)=3M_{P}^{2}\int_{\phi_{{\rm min}}}^{\phi_{{\rm max}}}d\phi V^{-1}(\phi)e^{-2v(\phi)}\times
\end{equation}
\begin{equation}
\times\int_{\phi}^{\phi_{{\rm max}}}d\phi'\, e^{2v(\phi')}\int_{\phi_{{\rm min}}}^{\phi'}d\phi''\, P(\phi,N_{{\rm i}}).
\end{equation}

Finally, the curvature perturbation correlation functions are related
to the moments $Q_{n}(\phi)$ according to the prescription
\begin{equation}
\langle\zeta^{n}(N_{{\rm total}})\rangle=n\int_{\phi_{{\rm min}}}^{\phi_{{\rm max}}}d\phi\, Q_{n-1}(\phi)-\left(\int_{\phi_{\rm min}}^{\phi_{{\rm max}}}d\phi\, Q_{0}(\phi)\right)^{n}.
\end{equation}

As we see, all one-point correlation functions of $N_{{\rm total}}$
and $\zeta$ (computed at the end of inflationary stage) are finite for
all physically interesting chaotic inflationary models. The final
answer does depend on the initial conditions for eternal inflation
due to the dependence of the zero moment $Q_{0}$ on the initial distribution
function $P(\phi,N_{i})$. For example, one has
\begin{equation}
\langle N_{{\rm total}}^{n}\rangle=C_{n}\frac{M_{P}^{2n}}{m^{2n}},
\end{equation}
\begin{equation}
\langle\zeta^{n}(N_{{\rm total}})\rangle=D_{n}\frac{M_{P}^{2n}}{m^{2n}}
\end{equation}
for the ${m^{2}\phi^{2}}$ chaotic inflation model,
where $C_{n},\, D_{n} = C_n - C_0^n\sim{\cal O}(1)$ are numerical constants depending
on the initial conditions. By virtue of the definition (\ref{eq:zetatotal}), $D_{1}=0$.

These large non-gaussianities however never become observable for
any given observer within a given Hubble patch, and we now turn to
a discussion of the observable quantities.

\section{The two-point \& three-point functions of the curvature perturbation at one loop}

In the previous section we demonstrated that the one-point functions
of the scalar field and the curvature perturbation are finite when
the effects of eternal inflation are taken into account. Thus, there
are no real infinities in the theory. Furthermore, the scales
alluded to in the previous section are much larger than any
currently observable scale. Then the question arises: How do the
predictions at scales $L$ translate into predictions at scales
$M < L$? We address the issue of relating two different scales in the present section using the
prescriptions of the $\Delta N$ formalism \cite{DeltaN}.

According to the $\Delta N$ formalism, the perturbative expansion
for the correlator of the curvature perturbation to one loop is given by \cite{BKSW}
    \beq \label{bare}
    \langle \zeta(\vc{x}_1)\zeta(\vc{x}_2)\rangle =\left(N_A N^A + N_A N^{AB}_B\langle\phi^2\rangle^{(l)}_{(L)} \right)
    G_{12} +\frac{1}{2}N_{AB}N^{AB}G^2_{12}\ ,
    \eeq
where $G_{ij}\equiv G(r_{ij})= G(|\vc{x}_i-\vc{x}_j|)=\langle\phi(\vc{x}_i)\phi(\vc{x}_j)\rangle$.
The parameters $L$ and $l$ in the symbol $\langle\phi^2\rangle^{(l)}_{(L)}$ are the infrared and ultraviolet cutoffs, respectively,
applied to all quantities in equation (\ref{bare}).
The 3-point function to one loop is given by
    \bea\label{3pt}
    \langle\zeta(\vc{x}_1)\zeta(\vc{x}_2)\zeta(\vc{x}_3)\rangle &=&
    \left[N_{AB}N^AN^B+\left(\frac{1}{2}N^A_{ABC}N^BN^C+N^A_{AB}N^{BC}N_C\right)\langle\phi^2\rangle^{(l)}_{(L)}\right] \left(G_{12}G_{13}+G_{21}G_{23}+G_{31}G_{32}\right)  \nonumber \\
     &+&\frac{1}{2}N_{ABC}N^{AB}N^C\Big(\left(G_{12}+G_{13}\right)G_{12}G_{13}+\left(G_{21}+G_{23}\right)G_{21}G_{23} +\left(G_{31}+G_{32}\right)G_{31}G_{32}\Big) \nonumber \\
    &+& N_{AB}N^{BC}N^A_C \,\,G_{12}G_{13}G_{23}\ ,
    \eea
where both of the above formulae assume that
    \beq
    \langle\phi^A(\vc{x}_i)\phi^B(\vc{x}_j)\rangle = \delta^{AB}G_{ij}\, .
    \eeq
Thus, $G(r)$ is the correlation function of a generic light field in quasi de-Sitter. The exact scalar field
correlator $G(r)$ is supposed to be known up to a superlarge scale $L$ and is presumably
well defined -- see previous section for its coincident limit. According to the $\Delta N$ prescription,
all the derivatives of $N$ in (\ref{bare}), (\ref{3pt}) are to be evaluated at the value of the background
field corresponding to some time after the shortest scale of interest has left the horizon.
Note that the expansion (\ref{bare}) requires the existence of a well defined background evolution. Then,
the picture of stochastic inflation imposes that the largest field value  for which the
expansion (\ref{bare}) is applicable is given by the hypersurface of eternal inflation
(\ref{eq:EternalInflationBoundary}) with $\phi_{\rm EI}$. For larger field values, $\phi > \phi_{\rm EI}$,
the notion of the background field breaks down and so does the expansion (\ref{bare}). Spacetime above this scale
can no longer be described by a global FRW metric and the correlators $G(r)$ are no longer translationally invariant.
However, it is important to note that an observer will never
interact with the surface of eternal inflation since such an interaction
presupposes a universe which has thermalized as only in the thermalized patches will the horizon grow,
permitting such measurements. 

Consider now an observer located in some random position. Since the state of the field is translationally invariant,
the stochastic properties of the field will be described by the correlation function $G(r)$ for any such observer.
However, the field fluctuations will not be experimentally accessible beyond some length scale $M$ defining the causal
patch of the observer. Fluctuations with wavelengths longer than $M$ should appear as part of the background for this patch.
Since these fluctuations only contribute a constant to $G(r)$ for $r<M$, the correlator appropriate for describing measurements within the
observer's patch is
    \beq
    \tilde{G}(r)\equiv G(r)-\langle\phi^2\rangle^{(M)}_{(L)} \,.
    \eeq
In other words, an observer limited to measurements in a volume  $r < M$ will be able to measure $G(r)$ up to a constant. One has
    \beq\label{tildeG}
    \tilde{G}(r)=\int\limits_{r/L}^{r/M} \frac{dp}{p} \mathcal{P}_\phi(p)\left(\frac{\sin p}{p}-1\right) +\int\limits_{1/M}^{1/l}
    \frac{dk}{k} \mathcal{P}_\phi(k) \frac{\sin k r}{k r}\,\Rightarrow \tilde{G}(0) = \langle\phi^2\rangle^{(l)}_{(M)}\,,
    \eeq
where $\mathcal{P}_\phi(k)$ is the power-spectrum obtained from the fourier transform of $G(r)$. The first term in $\tilde{G}$ (\ref{tildeG})
gives a small finite correction to the correlation function which goes to zero as $r$ becomes much smaller than $M$. The minimal length scale
$l$ corresponds to the shortest cosmological scale of interest; it is known to be at least 60 e-folds smaller than the size of the causal
patch $M$ (or larger depending on the temperature at reheating).

Let us now express Eq. (\ref{bare}) in terms of $\tilde{G}(r)$, the correlator appropriate for the patch of the size $M$. Keeping only one loop terms, we have
    \beq\label{renorm1}
    \langle\zeta(\vc{x}_1) \zeta(\vc{x}_2)\rangle =\left(N_AN^A+\frac{1}{2}(N_AN^A)^B_B\langle\phi^2\rangle^{(M)}_{(L)}
    +N_AN^{AB}_B\langle\phi^2\rangle^{(l)}_{(M)}\right)\tilde{G}_{12}+\frac{1}{2}N_{AB}N^{AB}\tilde{G}^2_{12} + N_AN^A\langle\phi^2\rangle^{(M)}_{(L)}\,.
    \eeq
The last term is a constant, which cancels out when comparing differences of two correlators.
The coefficient of $\tilde{G}(r)$ now includes the term
    \beq
    \Big\langle{N^AN_A}\left(\bar{\phi}_L+\phi(\vc{x})^{(M)}_{(L)}\right)\Big\rangle
=N_AN^A\Big\arrowvert_{\bar{\phi}_L}+\frac{1}{2}(N_AN^A)^B_B\Big\arrowvert_{\bar{\phi}_L} \langle\phi^2\rangle^{(M)}_{(L)}\,,
    \eeq
where $\bar{\phi}_L$ denotes the background field appropriate for the scale $L$ and $\phi(\vc{x})^{(M)}_{(L)}$ are fluctuations with
wavelengths larger than $M$. The 3-point
function gives a similar result. Expressed in terms of $\tilde{G}(r)$ it takes the form
    \bea\label{renorm-3pt}
    \langle\zeta(\vc{x}_1)\zeta(\vc{x}_2)\zeta(\vc{x}_3)\rangle
    &=& \left[\Big\langle{N_{AB}N^AN^B}\left(\bar{\phi}_L+\phi(\vc{x})^{(M)}_{(L)}\right)\Big\rangle
    +\left(\frac{1}{2}N^A_{ABC}N^BN^C+N^A_{AB}N^{BC}N_C\right)\langle\phi^2\rangle^{(l)}_{(M)}\right]
    \left(\tilde{G}_{12}\tilde{G}_{13}+\ldots\right) \nonumber \\
    &+&\frac{1}{2}N_{ABC}N^{AB}N^C\Big(\left(\tilde{G}_{12}+\tilde{G}_{13}\right)\tilde{G}_{12}\tilde{G}_{13}
    +\ldots \Big) + N_{AB}N^{BC}N^A_C \,\,\tilde{G}_{12}\tilde{G}_{13}\tilde{G}_{23}\nonumber \\
    &+& 2N_{AB}N^AN^B\langle\phi^2\rangle_{(L)}^{(M)}\left(\tilde{G}_{12}+\tilde{G}_{13}+\tilde{G}_{23}\right)+{\rm const}\ ,
    \eea
where we again have retained only terms up to one loop. We thus see that the result of
$L \rightarrow M$ is the replacement of the tree level coefficients by their averages,
    \beq\label{average}
     {\mathcal{A}}_n(M) = \Big\langle \mathcal{A}_n\left(\bar{\phi}_L+\phi(\vc{x})^{(M)}_{(L)}\right)\Big\rangle\,,
     \eeq
where $\mathcal{A}_1 = N^AN_A\,,\,\, \mathcal{A}_2 = N_{AB}N^AN^B$ etc, as well as the appearance of
a disconnected part which for the two point correlator is simply a constant. Otherwise, the functional form of the expansion
as given by the rules of \cite{BKSW} remains the same with all integrals now truncated at the new IR cutoff $M$.
The same should hold for all coefficients of all correlators in the $\Delta N$ expansion and therefore we write
    \beq
    \label{renorm-2}
    \langle\zeta(\vc{x}_1) \zeta(\vc{x}_2)\rangle_{M} =\left(\langle{N_AN^A}\rangle
    +\langle{N_AN^{AB}_B}\rangle\langle\phi^2\rangle^{(l)}_{(M)}\right)\tilde{G}(r)
    +\frac{1}{2}\langle{N_{AB}N^{AB}}\rangle\tilde{G}^2(r)+ \ldots\,,
    \eeq
and
    \bea\label{renorm-2-3pt}
    \langle\zeta(\vc{x}_1)\zeta(\vc{x}_2)\zeta(\vc{x}_3)\rangle_M
    &=& \left[\langle{N_{AB}N^AN^B}\rangle
    +\left(\frac{1}{2}\langle{N^A_{ABC}N^BN^C}\rangle+\langle{N^A_{AB}N^{BC}N_C}\rangle\right)\langle\phi^2\rangle^{(l)}_{(M)}\right]
    \left(\tilde{G}_{12}\tilde{G}_{13}+\ldots\right) \nonumber \\
    &+&\frac{1}{2}\langle{N_{ABC}N^{AB}N^C}\rangle\Big(\left(\tilde{G}_{12}+\tilde{G}_{13}\right)\tilde{G}_{12}\tilde{G}_{13}
    +\ldots \Big) + \langle{N_{AB}N^{BC}N^A_C}\rangle \,\,\tilde{G}_{12}\tilde{G}_{13}\tilde{G}_{23}\nonumber \\
    &+& 2\langle{N_{AB}N^AN^B}\rangle\langle\phi^2\rangle_{(L)}^{(M)}\left(\tilde{G}_{12}+\tilde{G}_{13}+\tilde{G}_{23}\right)+\ldots \,.
    \eea
Thus, the coefficients appropriate for the cutoff $M$ can be calculated
by replacing $\bar{\phi}_L \rightarrow \bar{\phi}_L+\phi(\vc{x})^{(M)}_{(L)}$  and averaging over all
field fluctuations $\phi(\vc{x})^{(M)}_{(L)}$ with wavelengths longer than $M$. This appears to be the generic form for the
$\Delta N$ expansion when used in patches smaller than $L_{EI}$.

Equation (\ref{average}) could also be interpreted in the following way: for an observer in the patch of the size $M$
making observations at $r<M$, the quantity $\bar{\phi}_L+\phi(\vc{x})^{(M)}_{(L)}$ is a slowly varying function,
indistinguishable from a background field. Of course, this background differs according to the location of the
patch in the larger volume of size $M$. Thus, the average in equation (\ref{average}) can be seen as an average over the
position of the patch of the size $M$ in the larger volume or, equivalently, an average over all possible ``backgrounds'' seen by
observers residing in $M$ \cite{Lyth1, bartoloetal}. Of course, this quantity is not (necessarily) related to observable quantities
of interest to any particular observer in a specific patch, in agreement with the results of \cite{Lyth1, bartoloetal}.

The above considerations might seem to suggest that the only appropriate infrared cutoff for observers like us
is the current horizon $L=1/H_0$ \cite{Lyth1}. Defining the theory at $L= 1/H_0$, seems to avoid the need of using averages in the
$\Delta N$ expansion. However, framed in this way, the choice of the IR cutoff
seems dictated by convenience rather than the theory itself and is hence ad hoc, designed to hide
the long wavelength fluctuations. However, fluctuations
on scales larger than our current horizon presumably exist and we ``simply'' have to wait a few billion years to
access them. In the future we will be forced to calculate using the cutoff appropriate for that time
and thus obtain a different result for our local curvature perturbation. Therefore, \emph{using a larger box should not introduce any
theoretical uncertainty for predictions of the local curvature perturbation field}.

Equation (\ref{renorm1}) allows us to deal with a change in the IR cutoff. We can define the coefficients in the expansion (\ref{bare})
to be evaluated at a background field appropriate for our current patch and impose $1/H_0$ as an IR cutoff. If now we are
granted access to longer wavelengths, say by waiting long enough and repeating our observations, it is expected that we will infer
a \emph{different} background evolution since we will now be averaging over a different volume and more long wavelength
fluctuations will be visible. Thus, we must accordingly revise the inflationary history that we match to the post-inflationary
universe. This will of course lead to a variation in the coefficients of the $\Delta N$ expansion for the connected part of the correlators, given by
    \beq\label{running}
    M\frac{d \mathcal{A}_n}{d M}= -\frac{1}{2}\left(\mathcal{A}_n\right)^B_B\,\,\mathcal{P}_{\phi}(1/M)\,.
    \eeq
This running shows how the connected parts of the correlation functions of the curvature perturbation
change as the cutoff scale $M$ is increased. It is akin to the renormalization group flow in quantum
field theory, where the coupling constants run with the renormalization scale, and we will accordingly refer to M
as the renormalization scale for cosmological perturbations. We note, however, that at least for the $2$-point function (\ref{renorm1})
the renormalization simply amounts to a constant change in the numerical value of the connected part of the correlator (see (\ref{bare})).
The difference of the $2$-point functions, which is a measurable quantity, thus remains invariant under the change of the cutoff scale
   \beq
   \label{diff2point}
   \langle\zeta(\vc{x}_1) \zeta(\vc{x}_2)\rangle-\langle\zeta(\vc{x}_1) \zeta(\vc{x}_3)\rangle=\langle\zeta(\vc{x}_1) \zeta(\vc{x}_2)\rangle_M-
   \langle\zeta(\vc{x}_1) \zeta(\vc{x}_3)\rangle_M\ .
   \eeq

Equation (\ref{running}) admits the following interpretation, at least in the case
of single field inflation. The coefficients $\mathcal{A}_n$ are functions of the derivatives
of the number of e-folds $N(\phi)$ w.r.t. $\phi$: $\mathcal{A}_n (N', N'', \ldots,N^{(n)})$. We assume that from (\ref{running}) one can derive
a consistent set of equations for the running of the derivatives of the form
    \beq\label{running-derivs}
    \frac{d N^{(n)}}{ d\ln M} = \mathcal{F}(N', N'', \ldots,N^{(n)}).
    \eeq
We see that as M is changed, the values of the derivatives $N^{(n)}(\phi)$, and thus the function $N(\phi)$ itself, are changed according to (\ref{running-derivs}).
Furthermore, equation (\ref{running-derivs}) takes a particularly simple form for inflationary models with a monomial potential $V(\phi)=\kappa \phi^n$.
Indeed, for such models
    \beq
    N'=-\frac{1}{n} \frac{8\pi}{M_{\rm P}^2} \phi~,~~ N''=-\frac{1}{n} \frac{8\pi}{M_{\rm P}^2}~,
    \eeq
and eqs (\ref{running-derivs}) simply become
    \beq
    \frac{d N'}{d\ln M} = -{1 \over 2}\frac{N''^2}{(N')}\mathcal{P}(1/M)\,.
    \eeq
We illustrate this point with a simple example in the next section
and refer to future work for further investigation.

\section{A concrete example: free field}

In this section, we consider two point correlators of the curvature perturbation
in single field inflation with quadratic potential $V=\frac{1}{2}m^2\phi^2$, although the following discussion would proceed
practically unchanged for any monomial potential.  We have
  \beq
  \label{nprimes}
  N'=-\frac{4\pi}{M_{\rm P}^2}\bar{\phi}~,~~N''=-\frac{4 \pi}{M_{\rm P}^2}\ ,
  \eeq
and all the higher order derivatives vanish.
The two point correlator (\ref{bare})
is thus written to all orders in loops as
  \beq
  \label{freetwo}
  \langle \zeta(\vc{x}_1)\zeta(\vc{x}_2)\rangle =\left(N'\right)^2
    G_{12} +\frac{1}{2}(N'')^2G^2_{12}\ .
  \eeq

According to (\ref{running-derivs}), the value of $N'$ will change as the cutoff is increased.
We parameterize the variation as
    \beq
    N'=-\frac{4\pi}{M_{\rm P}^2}\bar{\phi}_\star+ \beta(M)~,~~ \beta(1/H_0)=0~,
    \eeq
where we have assumed that the original cutoff has been set at our present horizon $M_0 =1/H_0$, such that expressions
(\ref{nprimes}) hold for this cutoff, and $\bar{\phi}_\star$ denotes the background field appropriate
for this cutoff. We thus have
    \beq
    \frac{d \beta}{d\ln M}  = \frac{1}{2}\left(\frac{4\pi}{M_{\rm P}^2}\right)^2 \frac{\mathcal{P}(1/M)}{\frac{4\pi}{M_{\rm P}^2}\bar{\phi}_\star+ \beta(M)}\,,
    \eeq
from which we obtain
    \beq
    \beta(M)=\frac{4\pi}{M_{\rm P}^2}\bar{\phi}_\star\left(1-\sqrt{1-\frac{\mathcal{P}}{2\bar{\phi}_\star^2}\ln H_0 M}\right)
    \eeq
where we have neglected the scale dependence of $\mathcal{P}$. Thus, with the new cutoff $M$, $N'$ reads
     \beq
     N'=-\frac{4\pi}{M_{\rm P}^2}\bar{\phi}_\star\sqrt{1-\frac{\mathcal{P}}{2\bar{\phi}_\star^2}\ln H_0 M}
     \equiv -\frac{4\pi}{M_{\rm P}^2}\bar{\phi}_M.
     \eeq
In this case, changing M simply amounts to evaluating expressions (\ref{nprimes}) at a \emph{new}
field value, shifted by
    \beq\label{shift}
    \bar{\phi} \rightarrow \bar{\phi}\sqrt{1-\frac{\mathcal{P}}{2\bar{\phi}^2}\ln H_0 M} \,.
    \eeq
Note that the form of the theory remains unchanged, ie still described by a quadratic potential. This is a general feature of all
monomial potentials.


So far, we have shown that one is free to choose the IR cutoff for inflationary perturbations, provided appropriate
shifts in the parameters of the theory are performed, see (\ref{shift}). It is also interesting to note
that for a free field it is simple to construct manifestly infrared
finite expressions containing the one-point correlators of the curvature perturbation. The expression for the two point
correlator (\ref{freetwo}) can be inverted to yield
  \beq
  \label{gee}
  G_{12}=\bar{\phi}^2\left(-1+\sqrt{1+(8\pi^2)^{-1}(M_{\rm P}/\bar{\phi})^4\langle\zeta(\vc{x}_1)\zeta(\vc{x}_2)\rangle}~\right)\ ,
  \eeq
where we have neglected the other unphysical root that would yield perturbations of the same order as the background field $G(r)\sim\bar{\phi}^{2}$.
The correlator $G(r)$ in (\ref{gee}) of course depends logarithmically on the IR cutoff but the difference
  \beq
  \label{diff}
  G(r)-G(r')=\int\limits_{1/L}^{1/l}dk\frac{\mc{P}_{\phi}(k)}{k}\left(\frac{{\rm sin}(kr)}{kr}-\frac{{\rm sin}(kr')}{kr'}\right)
  =\int\limits_{1/L}^{1/l}dk\frac{\mc{P}_{\phi}(k)}{k}\left(\frac{k}{6}(r'^2-r^2)+\ldots\right)
  \eeq
contains only negative powers of $L$ and is thus finite even in the limit $L\rightarrow\infty$. Thus we can construct a manifestly infrared finite
expression
  \beq
  \label{exdiff}
  G_{12}-G_{13}=\bar{\phi}^2\left(\sqrt{1+(8\pi^2)^{-1}(M_{\rm P}/\bar{\phi})^4\langle\zeta(\vc{x}_1)\zeta(\vc{x}_2)\rangle}-
  \sqrt{1+(8\pi^2)^{-1}(M_{\rm P}/\bar{\phi})^4\langle\zeta(\vc{x}_1)\zeta(\vc{x}_3)\rangle}~\right)\ .
  \eeq
By approximating the field correlator as
  \beq
  \label{apprgee}
  G(r)\approx-\mc{P}_{\phi}{\rm ln}\left(\frac{r}{L}\right)\ ,
  \eeq
one further obtains
  \beq
  \label{apprdiff}
  \sqrt{1+(8\pi^2)^{-1}(M_{\rm P}/\bar{\phi})^4\langle\zeta(\vc{x}_1)\zeta(\vc{x}_2)\rangle}-
  \sqrt{1+(8\pi^2)^{-1}(M_{\rm P}/\bar{\phi})^4\langle\zeta(\vc{x}_1)\zeta(\vc{x}_3)\rangle}=\frac{\mc{P}_{\phi}}
  {\bar{\phi}^2}{\rm ln}\left(\frac{r_{13}}{r_{12}}\right) ,
  \eeq
where the dependence on the cutoff has completely disappeared. This can be seen as a non-perturbative generalization of the first order result
(\ref{difference}), which is rederived by expanding (\ref{apprdiff}) to first order in
$(M_{\rm P}/\bar{\phi})^{4}\langle\zeta\zeta\rangle$. Since equation (\ref{apprdiff}) does not depend on the cutoff $L$ at all,
it remains unchanged in the renormalization prescription described above and is thus renormalization group
invariant to the precision of the approximation (\ref{apprgee}). Given a measurement of the two point correlator of
the curvature perturbation $\langle\zeta(\vc{x}_1)\zeta(\vc{x}_2)\rangle$ for one separation of points $r_{12}$,
equation (\ref{apprdiff}) can thus immediately be solved to find an unambiguous prediction for the correlator at any other separation $r_{13}$.

\section{Discussion}

The appearance of IR divergences in cosmological perturbation theory has been addressed by many authors with conclusions that do
not always coincide  \cite{Lyth1,Boubekeur:2005fj,Lyth:2006gd, bartoloetal, riotto-sloth}.
For example,
Lyth \cite{Lyth1} has suggested that the appropriate IR cutoff for inflationary calculations is a box slightly larger than the observable universe.
He argued that the use of a much larger box leads to problems since one would further need to average over the position of our observable
patch in this larger box and that this would introduce theoretical uncertainties. However, framed in this way, the choice of the IR cutoff seems
dictated by convenience rather than by the theory itself and is hence ad hoc. On the other hand, the authors of \cite{bartoloetal} have
pointed out that IR effects are indeed irrelevant for the observations but conceded that using a box larger than the horizon would require one
to average over the position of our patch leading to a prediction for an averaged, and therefore inherently statistical,
power spectrum. In contrast, we do not find any such theoretical uncertainty, nor
is there any need to average over the possible embeddings of our observable universe into the box defined by the hypersurface of
eternal inflation, since the field correlator is manifestly translation invariant. This due to the fact that
the box itself is defined by the end of inflation, i.e. each point has experienced the same fixed number of e-folds.

Using the framework of stochastic inflation, we first showed that for slow roll inflation the fluctuations are in fact finite to all scales up to Planckian
energy density\footnote{While this paper was nearing completion, \cite{riotto-sloth} appeared on the arXiv in which stochastic inflation
was used to demonstrate that the one-point function is finite for a special case of $\lambda\phi^4$ theory in pure de-Sitter.}. However,
fluctuations on such scales are unobservable, and only regions which have thermalized are interesting observationally. We then
related the predictions for the curvature perturbation on the largest possible thermalized scales to observations performed in much smaller regions
by an appropriate redefinition of the coefficients in the $\Delta N$ expansion.
As we have discussed, this procedure can be carried out to the case of 3-point functions, and we conjecture that this is so for any n-point correlator.

Our findings show that it is indeed permissible to use our horizon as a cutoff but that there is no uncertainty related to using a larger box.
In perturbation theory the difference between the two choices is an additive constant plus a redefinition of the background at which the coefficients of
the $\Delta N$ expansion (\ref{bare}) should be evaluated. In a sense, the size of the box $M$ represents simply the
renormalization point, the redefinitions a renormalization prescription. The largest possible box is defined by the requirement that
the field has dropped below the eternal inflation threshold (see Eq. (\ref{eq:EternalInflationBoundary})) and has therefore thermalized.
Since $G(r)$ is translationally invariant, any observer confined in a patch of size $M$ is able to probe it for $r<M$ and obtain the same
answer regardless of where the patch is located.

Thus, in general the correlators of the curvature perturbation were seen to depend on the renormalization scale $M$.
Since the (perturbative) renormalization of the two-point correlator amounts to a constant change in the numerical value of
the connected part of the correlator, the
difference of the two-point functions actually remains invariant under the change of the cutoff scale.
As discussed in section V, in the free single field case it is straightforward to find also a non-perturbative generalization
of the first order result which is independent of the infrared cutoff $L$. As is apparent from Eq. (\ref{apprdiff}),
given a measurement of the correlation of
the curvature perturbation between the points $\vc{x}_1$ and $\vc{x}_2$, one can find an unambiguous and renormalization point
independent prediction for any  $\langle\zeta(\vc{x}_1)\zeta(\vc{x}_3)\rangle$. It would be very interesting to generalize this result
to multifield cases as well as to non-trivial potentials.

\vskip30pt

\centerline{\textbf{Acknowledgements}}
\vskip10pt
We thank Alexei Starobinsky for illuminating discussions. This work
was supported by the EU 6th Framework Marie Curie Research and
Training network "UniverseNet" (MRTN-CT-2006-035863) and partly by
Academy of Finland grant 114419. S.N. is supported by the GRASPANP
Graduate School.


\begin{thebibliography}{20}

\bibitem{WMAP3}
D.N. Spergel, et.al., Astrophys. J. Suppl. 170:377,2007.

\bibitem{DeltaN}
A.~A.~Starobinsky Pis.~Zh.~Eksp.~Teor. {\bf 42}, 124 (1985);
A.~A.~Starobinsky JETP Lett. {\bf 42}, 124 (1985);
M.~Sasaki and E.~D.~Stewart,
  Prog.\ Theor.\ Phys.\  {\bf 95}, 71 (1996)
  [arXiv:astro-ph/9507001];
 D.~H.~Lyth, K.~A.~Malik and M.~Sasaki,
  JCAP {\bf 0505}, 004 (2005)
  [arXiv:astro-ph/0411220];
D.~H.~Lyth and Y.~Rodriguez, Phys.~Rev.~Lett. {\bf 95}, 121302 (2005),
[arXiv:astro-ph/0504045].


\bibitem{Lyth1} D.H. Lyth, JCAP {\bf 12}, 016 (2007), [arXiv:0707.0361
[astro-ph]].

\bibitem{Boubekeur:2005fj}
  L.~Boubekeur and D.~H.~Lyth,
  Phys.\ Rev.\  D {\bf 73} (2006) 021301
  [arXiv:astro-ph/0504046].


\bibitem{Lyth:2006gd}
  D.~H.~Lyth,
  JCAP {\bf 0606} (2006) 015
  [arXiv:astro-ph/0602285].

\bibitem{bartoloetal} N. Bartolo et. al., arXiv:0711.4263 [astro-ph].

\bibitem{riotto-sloth} A. Riotto and M.S. Sloth, arXiv:0801.1845 [hep-ph].

\bibitem{StarobinskyStochasticInflation}A.A. Starobinsky, in \emph{{}``Field
Theory, Quantum Gravity and Strings''}, eds. H.J. de Vega and N.
Sanchez, Lecture Notes in Physics (Springer-Verlag) \noun{246, 107}
(1986).

\bibitem{StarobinskyYokoyama}Y.~Nambu and M.~Sasaki, Phys.\ Lett.\  B {\bf 219}, 240 (1989); A.A. Starobinsky and J. Yokoyama, Phys.
Rev. D \textbf{50}, 6357 (1994) {[}arXiv:astro-ph/9407016].

\bibitem{AlStarOthers} S.~J.~Rey, Nucl.\ Phys.\  B {\bf 284}, 706 (1987); K.~i.~Nakao, Y.~Nambu and M.~Sasaki, Prog.\ Theor.\ Phys.\  {\bf 80}, 1041 (1988).

\bibitem{SalopekBondOthers}D.~S.~Salopek and J.~R.~Bond, Phys.\ Rev.\  D {\bf 43}, 1005 (1991);
S.~Mollerach, S.~Matarrese, A.~Ortolan and F.~Lucchin, Phys.\ Rev.\  D {\bf 44}, 1670 (1991).

\bibitem{curvaton}
K.~Enqvist and M.~S.~Sloth,
Nucl.\ Phys.\ B {\bf 626}, 395 (2002);
D.H. Lyth and D. Wands, Phys. Lett. B {\bf 524}, 5 (2002);
T.~Moroi and T.~Takahashi,
Phys.\ Lett.\ B {\bf 522}, 215 (2001)
[Erratum, ibid.\ B {\bf 539}, 303 (2002)].

\bibitem{LindeEternal}A. D. Linde, Phys. Lett. B \textbf{175}, 395
(1986); A. D. Linde, Mod. Phys. Lett. A \textbf{1},
81 (1986).

\bibitem{Woodard}R. P. Woodard, Nucl. Phys. Proc. Suppl. \textbf{148},
108 (2005) {[}arXiv:astro-ph/0502556].

\bibitem{LindeWaveFunction}A. D. Linde, Sov. Phys. JETP \textbf{60},
211 (1984); A. D. Linde, Lett. Nuovo Cimento \textbf{39},
401 (1984).

\bibitem{VilenkinWaveFunction}A. Vilenkin, Phys. Rev. D \textbf{30},
509 (1984); A. Vilenkin, Phys. Rev. D \textbf{33},
3560 (1986).

\bibitem{VilenkinSingularities}A. Borde and A. Vilenkin, Phys. Rev.
Lett. \textbf{72}, 3305 (1994).

\bibitem{LindeBook}A.~D.~Linde, \emph{``Particle Physics and Inflationary Cosmology,''}
Chur, Switzerland: Harwood (1990) [arXiv:hep-th/0503203].

\bibitem{BKSW} C.~T.~Byrnes, K.~Koyama, M.~Sasaki and D.~Wands, JCAP {\bf 0711}, 027 (2007)
  [arXiv:0705.4096 [hep-th]].





\end{thebibliography}
\end{document}